\documentclass{article}
\usepackage{natbib,psfig,emulateapj}

\newcommand\hi{{\small H\thinspace \footnotesize\sc\romannumeral 1 }}

\newcommand\oi{{\small O\thinspace \footnotesize\sc\romannumeral 1 }}
\newcommand\ovi{{\small O\thinspace \footnotesize\sc\romannumeral 6 }}

\newcommand\nh{\hbox{{$N_{\rm H}\,$}}}
\newcommand\nhgal{\hbox{{$N_{\rm H}^{\rm Gal}\,$}}}

\begin{document} 

\title{Oxygen Absorption in Cooling Flows}

\author{David A. Buote\altaffilmark{1}}

\affil{UCO/Lick Observatory, University of California at Santa Cruz,
Santa Cruz, CA 95064; buote@ucolick.org}

\altaffiltext{1}{Chandra Fellow}

\slugcomment{Accepted for Publication in The Astrophysical Journal
Letters} 

\begin{abstract}
The inhomogeneous cooling flow scenario predicts the existence of
large quantities of gas in massive elliptical galaxies, groups, and
clusters that have cooled and dropped out of the flow. Using spatially
resolved, deprojected X-ray spectra from the {\sl ROSAT} PSPC we have
detected strong absorption over energies $\sim 0.4-0.8$ keV intrinsic
to the central $\sim 1\arcmin$ of the galaxy, NGC 1399, the group, NGC
5044, and the cluster, A1795. These systems have amongst the largest
nearby cooling flows in their respective classes and low Galactic
columns. Since no excess absorption is indicated for energies below
$\sim 0.4$ keV the most reasonable model for the absorber is warm,
collisionally ionized gas with $T=10^{5-6}$ K where ionized states of
oxygen provide most of the absorption. Attributing the absorption only
to ionized gas reconciles the large columns of cold H and He inferred
from {\sl Einstein} and {\sl ASCA} with the lack of such columns
inferred from {\sl ROSAT}, and also is consistent with the negligible
atomic and molecular H inferred from \hi\, and CO observations of
cooling flows. The prediction of warm ionized gas as the product of
mass drop-out in these and other cooling flows can be verified by {\sl
Chandra}, {\sl XMM}, and {\sl ASTRO-E}.
\end{abstract}


\section{Introduction}
\label{intro}

The inhomogeneous cooling flow scenario (e.g., Fabian 1994) is often
invoked to interpret the X-ray observations of massive elliptical
galaxies, groups, and clusters.  The key prediction of this scenario
is the existence of large quantities of gas that have cooled out of
the hot phase and dropped out of the flow. The only evidence for large
amounts of mass drop-out arises from the excess soft X-ray absorption
from cold gas found for many cooling flows especially from spectral
analysis of {\sl Einstein} and {\sl ASCA} data (e.g., White et al
1991; Fabian et al 1994; Buote \& Fabian 1998; Buote 1999, 2000a).

This interpretation is highly controversial because for systems with
low Galactic columns no excess absorption from cold gas is ever found
with the {\sl ROSAT} PSPC which should be more sensitive because of
its softer bandpass, 0.1-2.4 keV (e.g., David et al 1994; Jones et al
1997; Briel \& Henry 1996). Furthermore, the large intrinsic columns
of cold H indicated by the {\sl Einstein} and {\sl ASCA} data are in
embarrassing disagreement with \hi\, and CO observations (e.g.,
Bregman, Hogg, \& Roberts 1992; O'Dea et al 1994).

We have found new evidence for absorption in the PSPC data during our
investigation of the radial metallicity profiles of several of the
brightest, nearby cooling flows (Buote 2000b, hereafter PAPER2).  In
this {\sl Letter} we present absorption profiles for a subset of these
cooling flows obtained from analysis of the deprojected PSPC spectra
and briefly compare these results to single-aperture {\sl ASCA}
spectra. We focus on the galaxy, NGC 1399, the group, NGC 5044, and
the cluster, A1795 because these systems have (1) low Galactic $N_{\rm
H}$ which facilitates analysis of any intrinsic absorption, (2) the
most significant excess columns measured from two-component spectral
models with {\sl ASCA}, and (3) the most clearly significant
differences in $N_{\rm H}$ obtained from PSPC data when energies below
$\sim 0.5$ keV are included/excluded from analysis.  The temperature
and metallicity profiles, as well as results for a few other systems
not satisfying all the above conditions, will be given in PAPER2 and
Buote (2000c, hereafter PAPER3).

\section{Spatially Resolved ROSAT Spectra}
\label{rosat}

We obtained archival {\sl ROSAT} PSPC observations of NGC 1399, NGC
5044, and A1795 and reduced the data as described in PAPER2. Here we
mention that particular attention was given to excluding any
significant fluctuations in the light curves since contamination from
solar emission would preferentially appear in the low-energy
channels. We also developed our own software to properly scale the
background spectra to the source positions.

For each object we extracted spectra in concentric circular annuli
located at the X-ray centroid such that for each annulus the width was
$\ge 1\arcmin$ and the background-subtracted counts was larger than
some value chosen to minimize uncertainties on the spectral parameters
for each system while maintaining as many annuli as possible. Data
with energies $\le0.2$ keV were excluded to insure that the PSF was
$<1\arcmin$ FWHM. For our on-axis sources $\sim 99\%$ of the PSF at
0.2 keV is contained within $R=1\arcmin$.

We deproject the data following McLaughlin (1999) who elaborates on
the original paper by Fabian et al (1981). The spectral fitting is
performed with XSPEC v10.0 using the MEKAL code to model the
(single-phase) hot plasma emission. We used the photoelectric
absorption cross sections of Baluci\'{n}ska-Church \& McCammon
(1992). Although Arabadjis \& Bregman (1999) point out that the He
cross section at 0.15 keV is in error by 13\%, since we analyze $E>
0.2$ keV we find that our fits do not change when using the Morrison
\& McCammon (1983) cross sections which have the correct He
value. Further details of our deprojection analysis are discussed in
PAPER2.

\subsection{Intrinsic Absorption from Cold Gas}
\label{cold}

No excess absorption from cold gas is found in cooling flows with the
PSPC when energies down to $\sim 0.1$-0.2 keV are included, whereas
significant absorption (similar to that obtained by {\sl ASCA}) is
found when the PSPC spectrum is restricted to energies above $\sim
0.5$ keV (Allen \& Fabian 1997; Buote 1999).  Presently, the only
viable explanation of why large excess absorbing columns are not
inferred when energies down to $\sim 0.1$-0.2 keV are included is that
the standard foreground screen model systematically underestimates the
true column intrinsic to the cooling flow (Allen \& Fabian 1997;
Sarazin, Wise, \& Markevitch 1998). Allen \& Fabian show in their
Figure 9 that a simulated PSPC cluster spectrum modified by both a
foreground column of $N_{\rm H}=10^{20}$ cm$^{-2}$ plus an intrinsic
column of $N_{\rm H}=10^{21}$ cm$^{-2}$ with covering fraction 0.5
will always yield a value similar to the foreground column if fitted
only with a standard foreground model. However, such two-component
models usually represent the projection of extended gas (with no
intrinsic absorption) with centrally concentrated, intrinsically
absorbed gas, and thus the suitability of these models can be tested
via deprojection.

\vspace*{0.2cm} \vbox{
\centerline{\psfig{figure=n1399_projected.ps,angle=-90,height=0.23\textheight}}
\figcaption[]{\label{fig.proj} \footnotesize {\sl ROSAT} PSPC spectrum
and projected model (see text). Statistical errors on the model are
similar to those shown for the data.}  }
\vspace*{0.2cm}

In Figure \ref{fig.proj} we display the PSPC spectrum for NGC 1399
within $R=1\arcmin$ (2D) and show the model representing the
projection of gas from $r>1\arcmin$ (3D) into the $R=1\arcmin$
aperture modified by a foreground Galactic absorber with solar
abundances.  The deprojection separates the foreground component
(i.e. the projected model in Fig \ref{fig.proj}) from the intrinsic
gas and shows that the intrinsic component cannot be absorbed very
differently from the outer gas since their spectral shapes below $\sim
0.5$ keV are so similar; e.g., for an intrinsic absorber with covering
factor of 0.5 we obtain a best-fit $\Delta N_{\rm H}=0$ and $\Delta
N_{\rm H}<N_{\rm H}^{\rm Gal}$ at $>90\%$ confidence. Since we obtain
analogous results for NGC 5044 and A1795, we conclude that the
deprojected PSPC spectra for each system rule out absorption by large
amounts of intrinsic cold gas (especially H and He) as a viable
explanation of the different excess absorbing columns obtained from
analyses in different X-ray bandpasses.

\subsection{Oxygen Edge}
\label{edge}

\begin{figure*}[t]

{\large\bf \hskip 2cm NGC 1399 \hskip 3.75cm NGC 5044 \hskip 4.2cm A1795}

\parbox{0.32\textwidth}{
\centerline{\psfig{figure=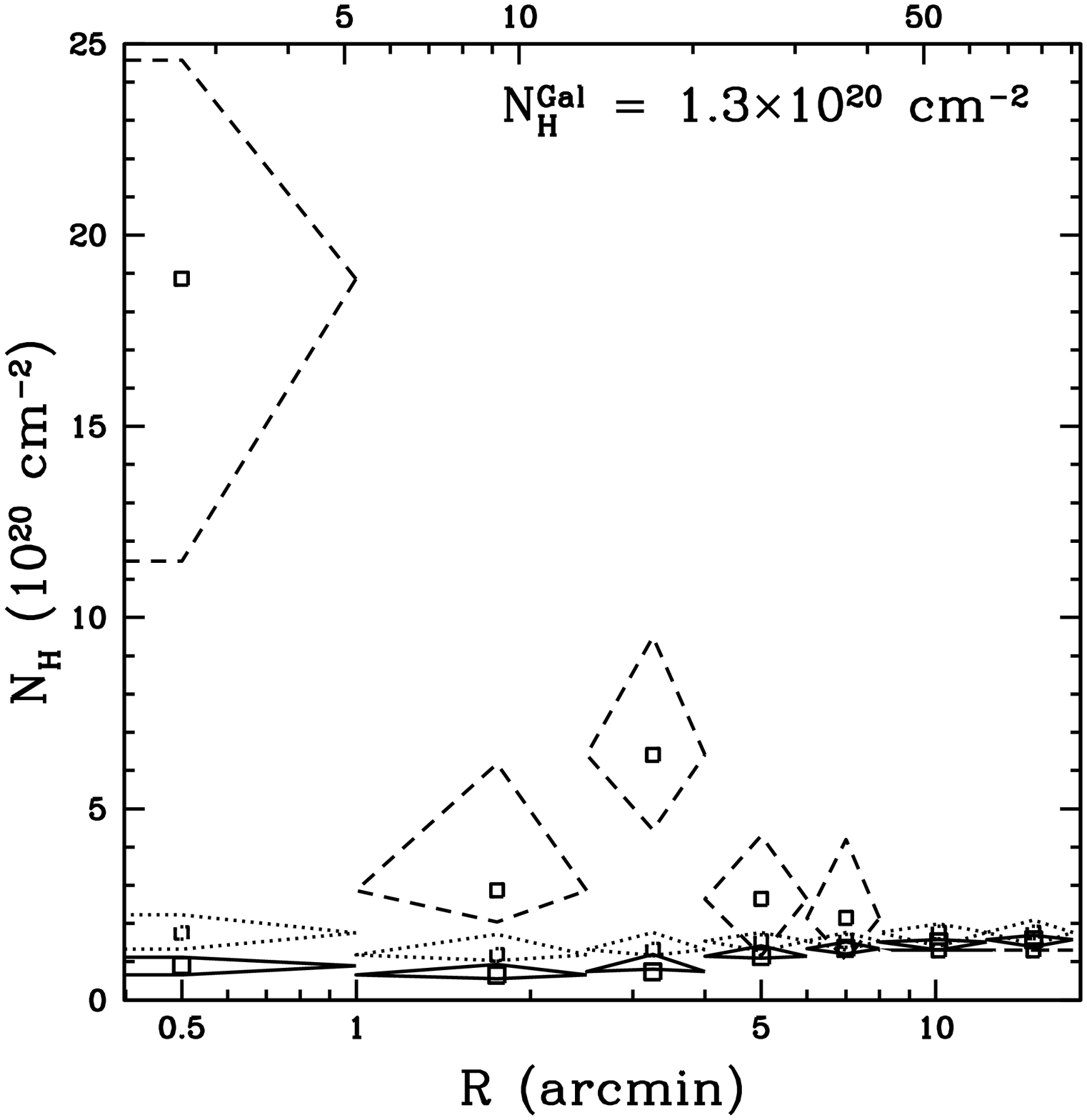,height=0.25\textheight}}
}
\parbox{0.32\textwidth}{
\centerline{\psfig{figure=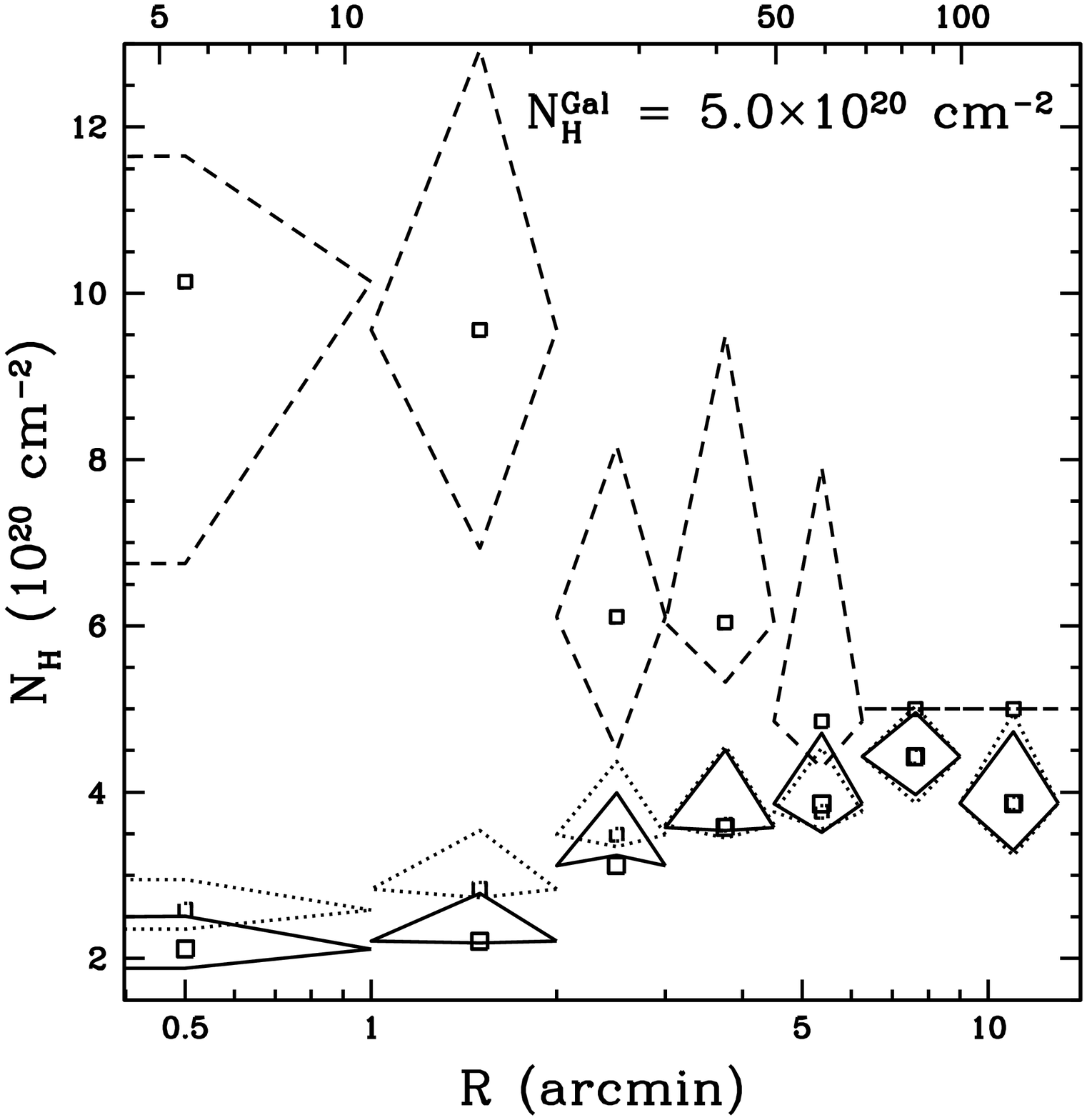,height=0.25\textheight}}
}
\parbox{0.32\textwidth}{
\centerline{\psfig{figure=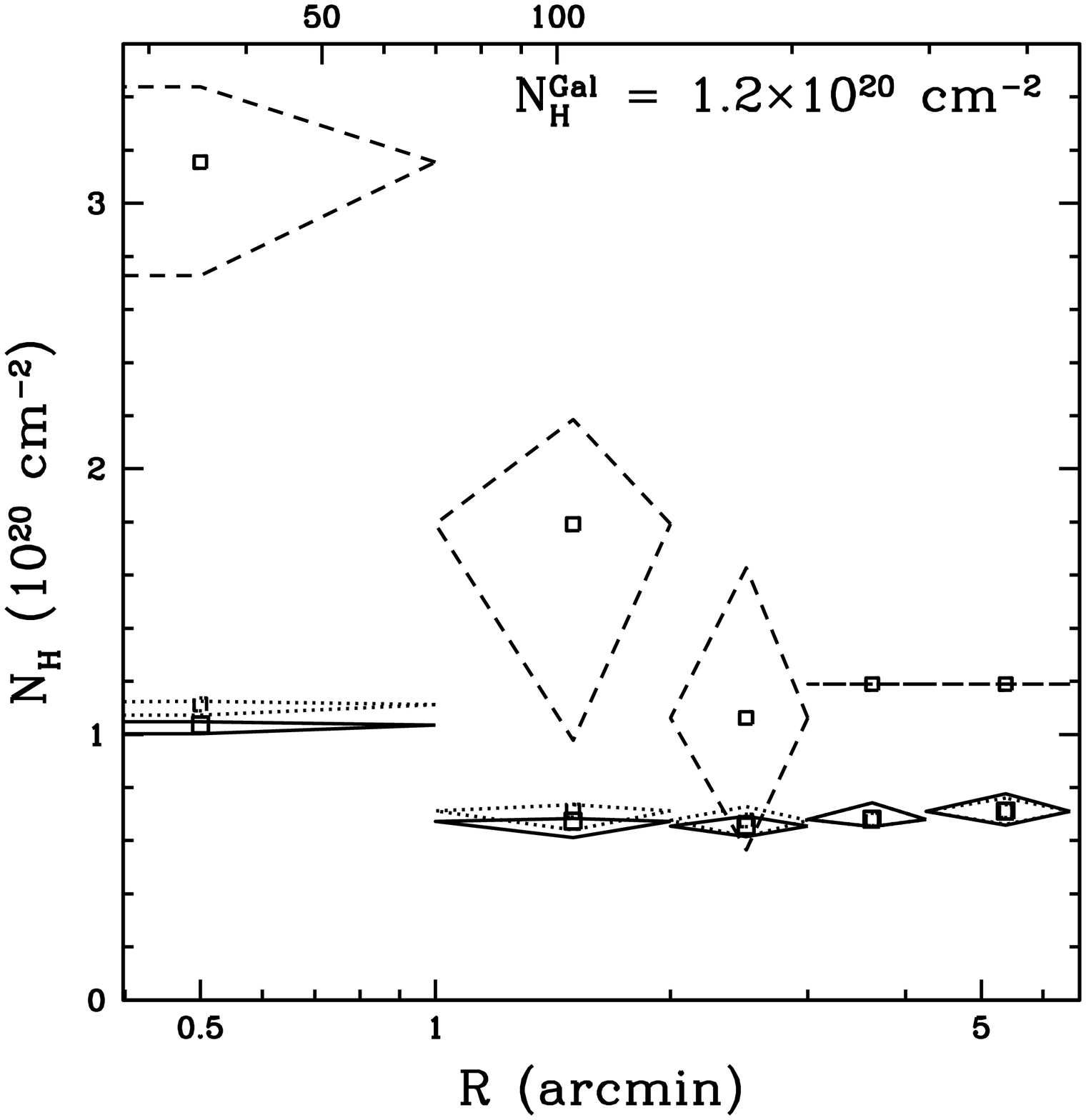,height=0.25\textheight}}
}
\vskip -0.1cm
\parbox{0.32\textwidth}{
\centerline{\psfig{figure=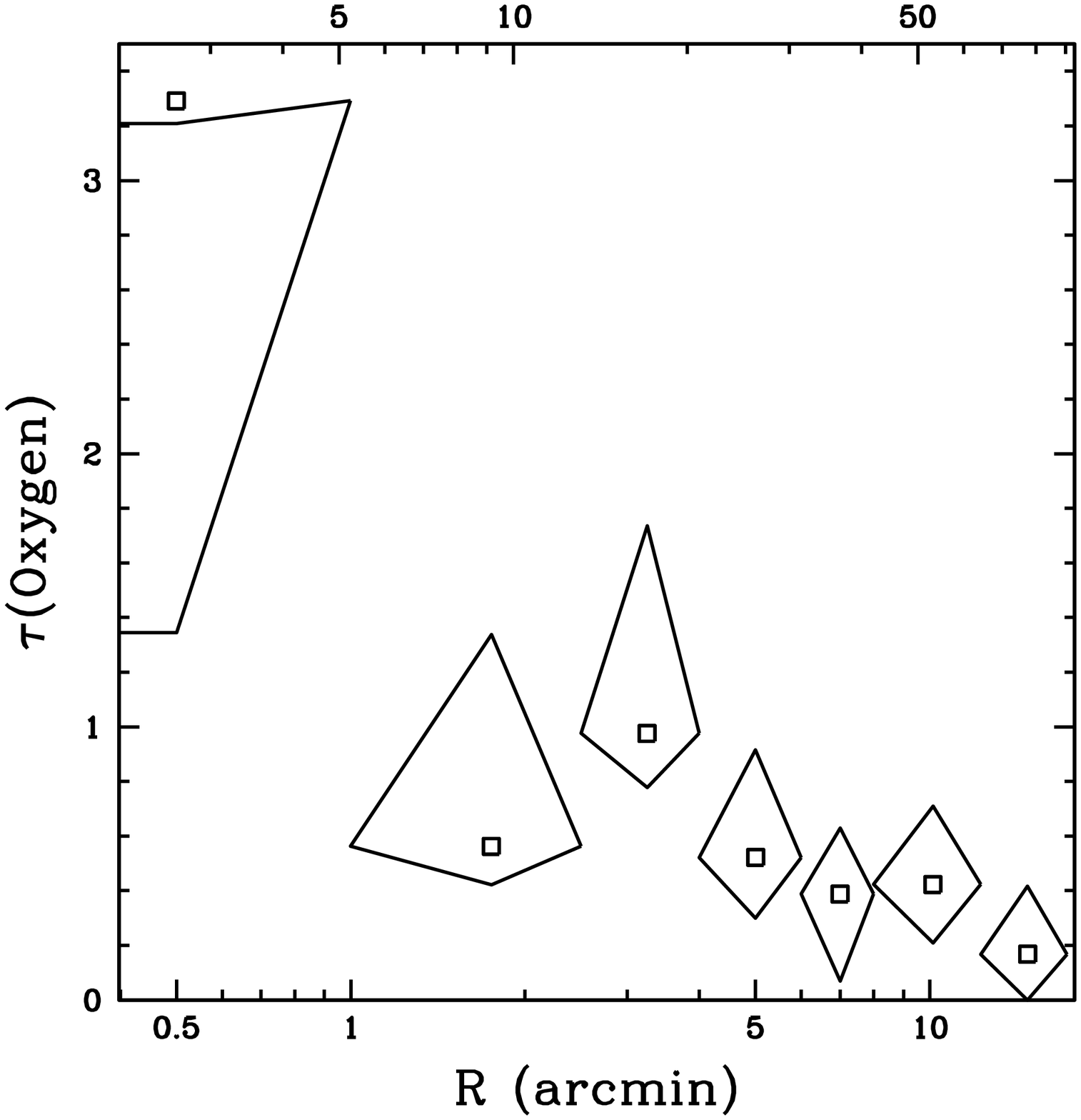,height=0.25\textheight}}
}
\parbox{0.32\textwidth}{
\centerline{\psfig{figure=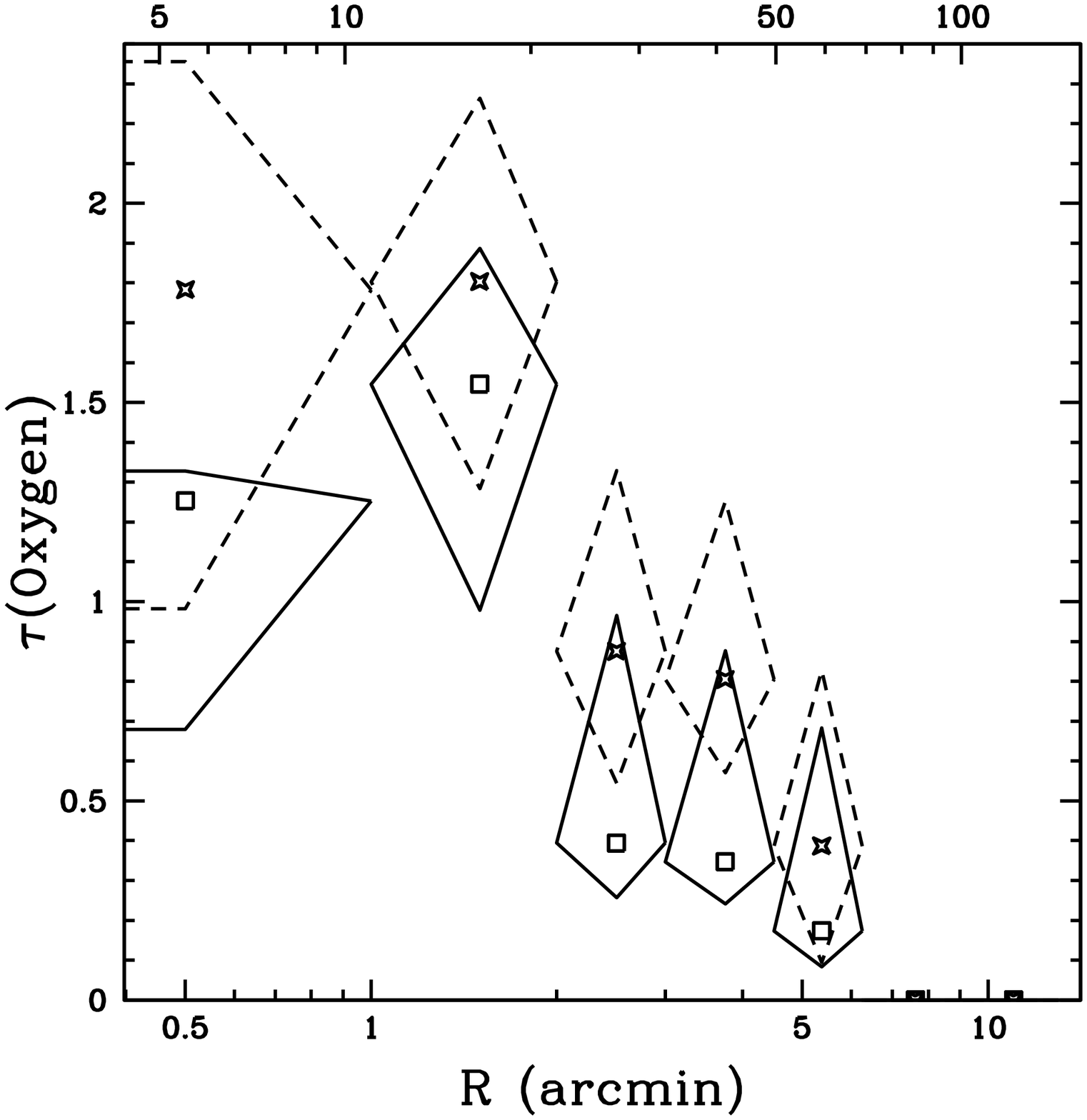,height=0.25\textheight}}
}
\parbox{0.32\textwidth}{
\centerline{\psfig{figure=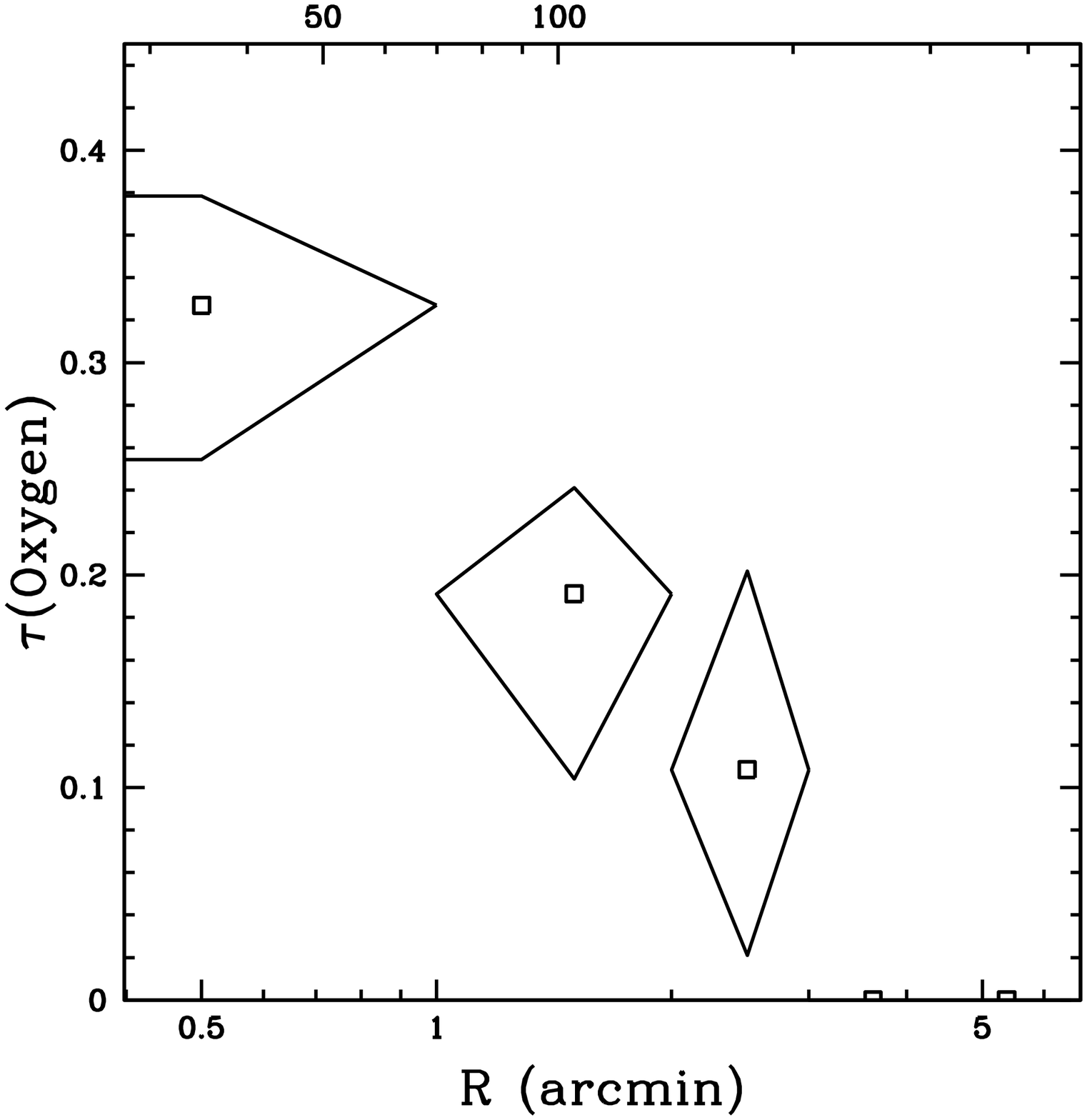,height=0.25\textheight}}
}
\caption{\label{fig.nh} (Upper panels) Radial column density profiles
and $1\sigma$ errors obtained from deprojection analysis of PSPC
spectra for fits over 0.2-2.2 keV (solid line), 0.5-2.2 keV (dashed),
and over 0.2-2.2 keV but including an O edge (dotted). (Lower panels)
Optical depth of 0.532 keV O edge (at redshift of object); dashed
diamonds for NGC 5044 represent case where $\nh\equiv\nhgal$. In the
outermost bins we set $\nh\equiv\nhgal$ for $E_{\rm min}=0.5$ keV and
$\tau\equiv 0$ because of weak constraints in some cases. Radial units
in kpc on top axis assume distances of 18, 38, and 240 Mpc
respectively for NGC 1399, 5044, and A1795.}

\end{figure*}

A solution to this problem becomes apparent upon examination of how
\nh varies with the lower energy limit in the centers of cooling
flows.  For example, in the central $1\arcmin$ of NGC 1399 we find
that $\nh\approx\nhgal$ when $E_{\rm min}\sim 0.2$-0.3 keV, though
when $E_{\rm min}\sim 0.4$ keV \nh increases to $\sim 2N_{\rm H}^{\rm
Gal}$. A dramatic change occurs for $E_{\rm min}\sim 0.5$ keV where
\nh increases to several times \nhgal, although for larger $E_{\rm
min}$ \nh remains nearly constant. In Figure \ref{fig.nh} we plot
$\nh(R)$ for $E_{\rm min}=0.2$, 0.5 keV for NGC 1399, 5044, and
A1795. The profiles radically differ for each $E_{\rm min}$: $\nh(R)
\sim \nhgal$ when $E_{\rm min}=0.2$ keV, but $\nh(R)$ increases
significantly at small $R$ when $E_{\rm min}=0.5$ keV such that
$N_{\rm H} > N_{\rm H}^{\rm Gal}$ at $>90\%$ confidence within
$R=1\arcmin$. (The uncertainties are computed from 100 Monte Carlo
realizations of the best-fitting models in each 2D annulus. Our simple
procedure for defining confidence limits (see PAPER2) occasionally
results in the best-fitting value lying outside the $1\sigma$ errors.)

Since $\nh(E_{\rm min})\approx const$ for $E_{\rm min} > 0.5$ keV the
portion of the spectrum responsible for the absorption must be $\sim
0.4$-0.7 keV considering the PSPC resolution. The most important
absorption (and emission) features over these energies are due to
oxygen. In Figure \ref{fig.nh} we also show the deprojection results
for $E_{\rm min}=0.2$ keV including an edge at 0.532 keV (rest frame)
corresponding to cold oxygen ({\small O\thinspace
\footnotesize\sc\romannumeral 1}). In this case $\nh(R)$ is very
similar to the previous fits with $E_{\rm min}=0.2$ keV and no edge
(i.e., similar to Galactic), and the edge optical depth profiles,
$\tau(R)$, closely resemble $\nh(R)$ when $E_{\rm min}=0.5$ keV.

The large optical depths at small radii and the improvements in the
fits when adding the edge are highly significant. In the $R=1\arcmin$
bins the 95\% confidence lower limits on $\tau$ are 0.97, 0.59, and
0.21 respectively for NGC 1399, 5044, and A1795.  Furthermore, 99 out
of 100 Monte Carlo simulations predict $\tau$ larger than 0.31, 0.10,
and 0.17 respectively; i.e., $\tau>0$ is highly significant for these
systems in the inner radial bin. The fits are improved substantially
when the edge is introduced in the central bin of each system -- see
Table \ref{tab.chi} where we list the values of $\chi^2$, degrees of
freedom (dof), and null hypothesis probability $(P)$. (Note that for
NGC 1399 the edge improves the fits significantly more than by letting
the oxygen abundance in the hot gas go to zero: i.e., $\chi^2$
improves only to 95.9 as opposed to 83.7 for the edge.)  We emphasize
that for the large values of $\tau\sim 1$ obtained in the central
regions the absorption from the edge affects a large energy range
comparable to the resolution of the PSPC over $\sim 0.5$-0.7 keV;
e.g., the \oi\, edge absorbs 25\% of the flux at 0.8 keV for $\tau=1$.

For most radii the constraints on the edge energy are not very precise
which is why we fixed the edge energy in our analysis. If we allow the
edge energy to be a free parameter in the central radial bin we obtain
$0.51^{+0.05}_{-0.05}$ keV, $0.51^{+0.09}_{-0.05}$ keV, and
$0.56^{+0.05}_{-0.05}$ keV (90\% confidence) for the edge energies of
NGC 1399, 5044, and A1795. These constraints are consistent with the
lower ionization states of oxygen but not edges from the highest
states {\small O\thinspace \footnotesize\sc\romannumeral
6-\footnotesize\sc\romannumeral 8}.  Due to the limited resolution we
can add additional edges to share the $\tau$ obtained for the \oi\,
edges, although even with a two-edge model a significant $\tau$ cannot
be obtained for edge energies above $\sim 0.65$ keV corresponding to
$\sim$\ovi.

(We mention that the models for NGC 5044 with $E_{\rm min}=0.2$ keV
not only imply $N_{\rm H} < N_{\rm H}^{\rm Gal}$ for small $R$ but
also predict very large metallicities that are inconsistent with the
{\sl ASCA} data. If \nh\, is fixed to \nhgal\, then consistent
metallicities are obtained as are $\tau(R)$ and $\chi^2$ improvements
very similar to the free \nh case -- see Figure \ref{fig.nh} and Table
\ref{tab.chi}; see also PAPER2 and PAPER3.)

{
\begin{table*}[t] \footnotesize
\begin{center}
\begin{minipage}{116mm}
\caption{$\chi^2$/dof/$P$ for Model Fits to Deprojected PSPC Spectra
within $r=1\arcmin$} 
\label{tab.chi}
\begin{tabular}{lcccc} \tableline\tableline
& NGC 1399 & \multicolumn{2}{c}{NGC 5044} & A1795\\ \\[-12pt]
& & Free $N_{\rm H}$ & $N_{\rm H}=N_{\rm H}^{\rm Gal}$\\ \tableline
no edge & 104.9/96/0.25 & 138.6/117/8.5e-2 & 191.2/118/2.3e-5 & 253.2/182/3.8e-4\\ 
w/ edge & 83.7/95/0.79  & 126.4/116/0.24 & 156.5/117/8.7e-3 & 212.0/181/5.7e-2\\  \tableline
\end{tabular}
%\tablecomments{ Stuff}
\end{minipage}
\end{center}
\end{table*}
}

\section{Single-Aperture ASCA Spectra}
\label{asca}

The intrinsic oxygen absorption indicated by the PSPC data is most
significant in the central $1\arcmin$ which is much smaller than the
width of the {\sl ASCA} PSF. In addition, since the {\sl ASCA} SIS is
limited to $E>0.5$ keV, and the efficiency near 0.5 keV is limited due
to instrumental oxygen absorption, it cannot be expected that {\sl
ASCA} can distinguish an oxygen edge from an absorber with solar
abundances. Nevertheless, we briefly summarize the results obtained
when adding an oxygen edge to the {\sl ASCA} data.

Previously we (Buote 1999) have fitted two-temperature models to the
accumulated {\sl ASCA} SIS and GIS data within $R\approx 5\arcmin$ of
NGC 1399 and 5044 and obtained $N_{\rm H}^{\rm c}=49_{-9}^{+6},
25_{-6}^{+5}\times 10^{20}$ cm$^{-2}$ respectively for the cooler
temperature components. Examination of Figure \ref{fig.nh} shows that
these columns are $\sim 2$ times the values in the central arcminute
and are consistent with the total columns within $R\approx 5\arcmin$
obtained from the PSPC for $E_{\rm min}=0.5$ keV. If instead we add an
\oi\, edge to the cooler {\sl ASCA} model components (while keeping
\nh\, fixed to Galactic on each component) we obtain (1) fits of
comparable quality (slightly better) to the original models and (2)
optical depths that agree with the cumulative values obtained from the
PSPC.

For A1795 we have re-analyzed the {\sl ASCA} data within $R\approx
4\arcmin$ and obtained results very similar to Fabian et al (1994) for
two-temperature models. As above, we obtain fits with an oxygen edge
supplying the excess absorption that are as good as those obtained
with an excess absorber having solar abundances. Unlike the
ellipticals and groups, for the cluster A1795 the cooler temperature
component contributes only $\sim 1/7$ to the emission measure, and
thus its fitted parameters are not nearly as well constrained as for
NGC 1399 and 5044; i.e., although the best-fitting $\tau=2.5$, the
90\% lower limit is 0.6 which is comparable to the total optical depth
inferred from the PSPC (Fig \ref{fig.nh}).

Hence, the oxygen edge provides as good or better description of the
excess absorption inferred from multitemperature models of {\sl ASCA}
data within the central few arcminutes of NGC 1399, 5044, and A1795 as
a cold absorber with solar abundances, and yields optical depths that
are consistent with those obtained with the PSPC data.

\section{Discussion}
\label{disc}

Using spatially resolved, deprojected X-ray spectra from the {\sl
ROSAT} PSPC we have detected strong oxygen absorption intrinsic to the
central $\sim 1\arcmin$ of the elliptical galaxy, NGC 1399, the group,
NGC 5044, and the cluster, A1795 which are amongst the largest nearby
cooling flows in their respective classes and have low Galactic
columns (e.g., Fabian et al 1994; Buote \& Fabian 1998; Buote 1999,
2000a). Modeling the oxygen absorption with an edge (rest frame
$E=0.532$ keV) produces the necessary absorption in both the PSPC and
{\sl ASCA} data for $E\ga 0.5$ keV without violating the PSPC
constraints over $0.2\sim 0.3$ keV for which no significant excess
absorption is indicated. This reconciles many reported discrepancies
between absorbing columns inferred from {\sl ROSAT} with those
obtained from {\sl ASCA} and other instruments with bandpasses above
$\sim 0.5$ keV. Moreover, there is no need for large absorbing columns
of cold H which are known to be very inconsistent with the negligible
atomic and molecular H measured in cooling flows (e.g., Bregman et al
1992; O'Dea et al 1994).

Since no excess absorption is detected over $0.2\sim 0.3$ keV, and
since the edge energies allow for ionized states of oxygen, the most
reasonable model for the absorber is warm, collisionally ionized gas
with $T=10^{5-6}$ K. At these temperatures the majority of the 
absorption arises from oxygen but with a sizeable contribution from
ionized carbon and nitrogen (see PAPER3).

Arnaud \& Mushotzky (1998) have reported the detection of an \oi\, edge
in the Perseus cluster using BBXRT data. Their cooling flow model
(which has stronger oxygen lines than two-temperature models) requires
an edge energy that is consistent only with cold oxygen, and thus warm
ionized gas could not be responsible for the absorption. However, it
was not shown that multiple edges from different ionization states are
inconsistent with the data, and thus we expect that a multi-edge model
would fit as well as a single \oi\, edge since the BBXRT has lower
energy resolution than {\sl ASCA}. (We do not find any evidence for
excess oxygen absorption in Perseus from the PSPC data probably
because of the large Galactic column, but $\tau\sim 2$ for the \oi\,
edge obtained by Arnaud \& Mushotzky is similar to what we found for
A1795 with {\sl ASCA}.)

The warm ionized gas implied for NGC 1399, 5044, and A1795 (and
possibly Perseus) could be the much sought-after mass that has dropped
out of the cooling flows. In PAPER3 we show that the absorber masses
indicated by the oxygen edge optical depths are consistent with the
mass expected to have been deposited by the cooling flow over the
lifetime of the system, while the emission expected from the warm gas
does not violate published optical and UV constraints. Understanding
the details of how warm ionized gas is maintained and supported in
such large quantities is a serious theoretical challenge.

Since detection of the warm ionized gas at other wavelengths is
difficult, e.g., background QSOs have not been found within
$R=1\arcmin$ of strong cooling flows (Miller, Bregman, \& Knezek
2000), confirmation will probably have to be realized in the X-ray
band.  Fortunately, the CCDs of the {\sl Chandra} and {\sl XMM}
missions have the combined spatial and spectral resolution and include
energies down to $\sim 0.1$ keV to verify our prediction of warm
ionized gas in these and other cooling flows. The $\sim 10$ eV
resolution of the {\sl ASTRO-E} XRS will place strong constraints on
the allowed ionization states of oxygen. Hence, data from these new
missions will elucidate the properties of the warm ionized gas and
thus the role of cooling flows in the formation and evolution of
ellipticals, groups, and clusters.

\acknowledgements

I thank M. Bolte and the referee for comments on the
manuscript. Support for this work was provided by NASA through Chandra
Fellowship grant PF8-10001 awarded by the Chandra Science Center,
which is operated by the Smithsonian Astrophysical Observatory for
NASA under contract NAS8-39073.

\end{document}